\documentclass[twocolumn,a4paper,superscriptaddress]{revtex4}
\usepackage{graphicx,amssymb}
\begin{document}
%My commands
\newcommand{\be}{\begin{equation}}
\newcommand{\ee}{\end{equation}}
\newcommand{\bq}{\begin{eqnarray}}
\newcommand{\eq}{\end{eqnarray}}
\newcommand{\bsq}{\begin{subequations}}
\newcommand{\esq}{\end{subequations}}
\newcommand{\bc}{\begin{center}}
\newcommand{\ec}{\end{center}}
\newcommand {\R}{{\mathcal R}}
\newcommand{\al}{\alpha}
\newcommand\lsim{\mathrel{\rlap{\lower4pt\hbox{\hskip1pt$\sim$}}
    \raise1pt\hbox{$<$}}}
\newcommand\gsim{\mathrel{\rlap{\lower4pt\hbox{\hskip1pt$\sim$}}
    \raise1pt\hbox{$>$}}}

\title{Assessing the viability of successful reconstruction of the dynamics of dark energy using varying fundamental couplings}

\author{P.P. Avelino}
\email[Electronic address: ]{ppavelin@fc.up.pt}
\affiliation{Centro de Astrof\'{\i}sica da Universidade do Porto, Rua das Estrelas, 4150-762 Porto, Portugal}
\affiliation{Departamento de F\'{\i}sica e Astronomia, Faculdade de Ci\^encias, Universidade do Porto, Rua do Campo Alegre 687, 4169-007 Porto, Portugal}
\author{L. Losano}
\email[Electronic address: ]{losano@fisica.ufpb.br}
\affiliation{Departamento de F\'{\i}sica, Universidade Federal da Para\'{\i}ba 58051-970 Jo\~ao Pessoa, Para\'{\i}ba, Brasil}
\affiliation{Centro de F\'{\i}sica do Porto, Rua do Campo Alegre 687, 4169-007 Porto, Portugal}
\affiliation{Departamento de F\'{\i}sica e Astronomia, Faculdade de Ci\^encias
da Universidade do Porto, Rua do Campo Alegre 687, 4169-007 Porto, Portugal}
\author{R. Menezes}
\email[Electronic address: ]{rmenezes@dce.ufpb.br}
\affiliation{Departamento de Ci\^encias Exatas, Universidade Federal da Para\'{\i}ba, 58297-000 Rio Tinto PB, Brazil}
\affiliation{Departamento de F\' \i sica, Universidade Federal de Campina Grande, 58109-970, Campina Grande, Para\'\i ba, Brazil}
\author{J.C.R.E. Oliveira} 
\affiliation{Centro de F\'{\i}sica do Porto, Rua do Campo Alegre 687, 4169-007 Porto, Portugal}
\affiliation{Departamento de Engenharia F\'{\i}sica da Faculdade de Engenharia da Universidade do Porto, Rua Dr. Roberto Frias, s/n, 4200-465 Porto, Portugal}
\email[Electronic address: ]{jespain@fe.up.pt}

\date{\today}
\begin{abstract}

We assess the viability of successful reconstruction of the evolution of the dark energy equation of state using varying fundamental couplings, such as the fine structure constant or the proton-to-electron mass ratio. We show that the same evolution of the dark energy equation of state parameter with cosmic time may be associated with arbitrary variations of the fundamental couplings. Various examples of models with the same (different) background evolution and different (the same) time variation of fundamental couplings are studied in the letter. Although we demonstrate that, for a broad family of models, it is possible to redefine the scalar field in such a way that its dynamics is that of a standard quintessence scalar field, in general such redefinition leads to the breakdown of the linear relation between the scalar field and the variation of fundamental couplings. This implies that the assumption of a linear coupling is not sufficient to guarantee a successful reconstruction of the dark energy dynamics and consequently additional model dependent assumptions about the scalar field responsible for the dark energy need to be made.

\end{abstract}
\pacs{}
\keywords{Cosmology; Dark energy}
\maketitle

\section{\label{intr}Introduction}

More than one decade ago type Ia supernovae observations suggested, for the first time, that the expansion of the universe is accelerating \cite{Perlmutter:1998np,Riess:1998cb}. Since then, increasingly precise cosmological observations \cite{Percival:2009xn,Komatsu:2010fb,Amanullah:2010vv} led to a well tested cosmological model presently dominated by an exotic dark energy form, violating the strong energy condition. In fact, if General Relativity is valid on large cosmological scales then dark energy \cite{Ratra:1987rm,Frieman:1995pm,Caldwell:1997ii,Ferreira:1997hj,ArmendarizPicon:2000dh,ArmendarizPicon:2000ah} provides the only convincing explanation for the observed acceleration of the universe. Understanding the nature of dark energy is therefore one of the most important challenges of modern cosmology with one of the primary goals being determining whether its energy density is constant or slowing varying (see \cite{Copeland:2006wr,Frieman:2008sn,Caldwell:2009ix,Silvestri:2009hh,Li:2011sd} for recent dark energy reviews). A fundamental problem associated to a cosmological constant is that its magnitude is constrained to be much smaller than particle physics predictions. On the other hand, it is also not clear if there is a deep physical reason which explains why it became the dominant component of the Universe just around the present day \cite{Barreira:2011qi}. 

An arguably better motivated alternative to the cosmological constant is the possibility that dark energy might be described by a dynamical scalar field. One important parameter characterizing dynamical dark energy is its equation of state, the ratio $w$ between the dark energy pressure and energy density. Constant $w$ models are unrealistic unless $w=-1$, which corresponds to the cosmological constant case \cite{Avelino:2009ze,Avelino:2011ey}. Hence, a measure of $w \neq -1$ at any redshift or redshift band should be indicative of dynamical dark energy. Considerable efforts are being put forward to constrain the dynamics of $w$ at low redshifts (see \cite{EditorialTeam:2011mu} for expected future developments with the Euclid mission) using type Ia supernova, galaxy clustering or weak lensing. These are indirect probes which rely on the impact of dark energy on the overall dynamics of the universe. However, dark energy is expected to become subdominant at early times and, consequently, it is not possible to strongly constrain its dynamics at high redshift using standard methods. 

Still, in realistic models, dynamical scalar fields may couple to other fields, possibly leading to measurable variations of nature's fundamental "constants" \cite{Carroll:1998zi}. The coupling between a quintessence field and fundamental couplings such as $\alpha$ or $\mu$ has been investigated by several authors \cite{Carroll:1998zi,Chiba:2001er,Wetterich:2002ic,Nunes:2003ff,Anchordoqui:2003ij,Copeland:2003cv,Avelino:2004hu,Parkinson,Doran:2004ek,Marra:2005yt,Avelino:2006gc,Avelino:2008dc,Avelino:2009fd,Dent:2009,Calabrese:2011nf,Amendola:2011qp}. The dynamics of $\alpha$ over the redshift range $z=0-10^{10}$  is severely constrained using using both cosmological and laboratory experiments (see \cite{Uzan:2010pm} for a recent review). At low redshifts laboratory experiments \cite{Peik:2006xy,Rosenband:2008} and the Oklo natural nuclear reactor \cite{Gould:2006,Petrov:2006,Onegin:2010kq}  provide very stringent limits on the time-variation of $\alpha$ and $\mu$, while at high redshift cosmic microwave background temperature and polarization anisotropies \cite{Avelino:2000ea,Avelino:2001nr,Martins:2003pe,Rocha:2003gc, Stefanescu:2007aa,Nakashima:2008cb} and light element abundances \cite{Bergstrom:1999wm,Avelino:2001nr,Nollett:2002da} constrain the value of $\alpha$ at $z \sim 10^{10}$ and $z \sim 10^3$ to be within a few percent of its present day value. Despite a few positive claims for a detection of a variation of the fine-structure constant $\alpha$ \cite{Webb:1998cq,Murphy:2006vs} or the proton-to-electron mass ratio $\mu$ \cite{Ivanchik:2005ws,Reinhold:2006zn} in the redshift range $z=1-4$, and the more recent claims for a significant spatial variation of $\alpha$ \cite{Webb:2011,King:2012}, there is presently no unambiguous  evidence for such variation (see, for example, \cite{Chand2004,Chand2007,Murphy2008} for some strong negative results). Nevertheless, it has been shown that varying couplings may be used to determine the evolution of the dark energy equation of state over a larger redshift range than standard methods, if a number of conditions are verified \cite{Avelino:2006gc,Avelino:2009fd}. These are: i) that the dark energy can be described by a standard quintessence field; ii) that the relation between the quintessence field and varying fundamental couplings  is linear; iii) that such variations are within reach of forthcoming experiments. In this letter we shall relax assumptions i) and ii) and consider more general k-essence models for dark energy, thus testing the robustness of the varying fundamental couplings method for the reconstruction of the evolution of the dark energy equation of state.

Throughout this letter we shall use units with $c=8\pi G/3=H_0=1$ and a metric signature $(+,-,-,-)$.

\section{Dynamics of varying couplings}

Consider a class of models described by the action
\begin{equation}\label{eq:L}
S=\int d^4x \, \sqrt{-g} \mathcal \, {\cal L} \, ,
\end{equation}
where ${\mathcal L}$ is given by
\begin{equation} 
{\cal L} = {\cal L}_\phi + {\cal L}_{\phi F} + {\cal L}_{\rm other}\, , 
\end{equation} 
and
\begin{eqnarray}
{\cal L}_\phi&=&{\cal L}_\phi(\phi,X)\,, \\
X&=&\frac{1}{2}\phi^{,\mu}  \phi_{,\mu} \,, \label{eq:kinetic_scalar1}\\
{\cal L}_{\phi F}&=& -\frac{1}{4}  B_F (\phi) F_{\mu \nu} F^{\mu \nu}\,, 
\end{eqnarray}
a comma represents a partial derivative, $B_F(\phi)$ is the gauge kinetic function, $F_{\mu \nu}$ are the components of the 
electromagnetic field tensor and ${\cal L}_{\rm other}$ is the Lagrangian of the other fields. The 
fine-structure constant is then given by 
\be 
\alpha(\phi)=\frac{\alpha_0}{B_F(\phi)} 
\label{gkfalpha} 
\ee 
and, at the present day, one has $B_F(0)=1$ (where the subscript `$0$' refers to the present time ). 

We shall also make the crucial assumption that the gauge kinetic function is a linear 
function of $\phi$ so that one has 
\be 
\frac{\Delta\alpha}{\alpha}\equiv \frac{\alpha-\alpha_0}{\alpha_0}=\beta \phi\,,
\label{gkfspec} 
\ee 
where $\beta$ is a constant and $\phi_0=0$. In the case of the standard quintessence model, the variations of the fine structure constant associated ${\cal L}_{\phi F}$ are very small given Equivalence Principle constraints \cite{Olive:2001vz}. For simplicity, we shall neglect the contribution of this term to the dynamics of $\phi$. This will not affect the conclusions of the letter.

The energy-momentum tensor associated with the scalar field $\phi$ may be written 
in a perfect fluid form
\begin{equation}\label{eq:fluid}
T^{\mu\nu}_\phi = (\rho_\phi + p_\phi) u^\mu u^\nu - p_\phi g^{\mu\nu} \,,
\end{equation}
by means of the following identifications
\begin{equation}\label{eq:new_identifications}
u_\mu = \frac{\phi_{,\mu}}{\sqrt{2X}} \,,  \quad \rho_\phi = 
2 X {\mathcal L}_{\phi,X} - {\mathcal L}_\phi \, ,\quad p_\phi =  {\mathcal L}_\phi\,,
\end{equation}
so that the equation of state parameter is given by
\begin{equation}
w_\phi=\frac{{\mathcal L}_{\phi}}{2X{\mathcal L}_{\phi,X}-{\mathcal L}_{\phi}}\,,
\end{equation}
and, if ${\mathcal L}_{\phi,X} \ne 0$, the sound speed squared is
\begin{equation}
\label{eq:cs2}
c_{s\phi}^2 \equiv \frac{p_{\phi,X}}{\rho_{\phi,X}}=\frac{{\mathcal L}_{\phi,X}}{{\mathcal L}_{\phi,X}+2X{\mathcal L}_{\phi,XX}}\,.
\end{equation}
In Eq.~(\ref {eq:fluid}), $u^\mu$ is the 4-velocity field describing the motion of the fluid (for timelike $\phi_{,\mu}$), while $\rho_\phi$ and $p_\phi$ are its proper energy density and pressure, respectively. The equation of motion for the scalar field is now
\begin{equation}
{\tilde g}^{\mu \nu} \phi_{;\mu\nu}={\mathcal L}_{,\phi}-2X{\mathcal L}_{\phi,X\phi}\,,
\end{equation}
where
\begin{equation}
{\tilde g}^{\mu \nu}={\mathcal L}_{\phi,X} g^{\mu \nu} + {\mathcal L}_{\phi,XX} \phi^{,\mu} \phi^{,\nu}\label{eq:dyn}\,.
\end{equation}

In this letter we consider a flat homogeneous and isotropic Friedmann-Robertson-Walker universe, permeated with minimally coupled matter and dark energy fluids. The dark energy is assumed to be described by the scalar field $\phi$. If $c_{s\phi}^2$ is sufficiently large then the spatial variations of $\phi$ may be neglected in Eq.~(\ref {eq:dyn}) \cite{Barrow:2005sv,Shaw:2005gt,Avelino:2005pw,Avelino:2008cu}. In this case the dynamics of the universe is described by
\begin{eqnarray}
H^2 = \rho_m+\rho_\phi=\rho_\phi+\Omega_{m0}e^{-3y}\,,\label{fried1}\\
\rho_\phi' = - 3(\rho_\phi+p_\phi)= -6 X {\mathcal L}_{\phi,X}\,,\label{consphi}
\end{eqnarray}
where the universe is assumed to be flat, the subscript `0' stands for the present time,  $H={\dot a}/a$ is the Hubble parameter with the dot representing a derivative with respect to physical time $t$, the time unit is chosen such that $H_0=1$, $\rho_m$ is the matter density, $\rho_\phi$ is the dark energy density, $p_\phi$ is the dark energy pressure, $a$ is the scale factor with $a_0=1$,  and a prime represents a derivative with respect to $y=\ln a$. We shall assume, for simplicity, that the parameters $w_{\phi 0}$, $\Omega_{m0}=\rho_{m0}$ are known a priori (in a flat universe filled with dark matter and dark energy $\Omega_{\phi 0}=\rho_{\phi0}=1-\Omega_{m0}$). This implies that the system constituted by Eqs. (\ref {fried1}) and (\ref {consphi}) may be solved if one knows the dependence of the right hand side of Eq. (\ref {consphi}) on $y$, $H$ and $\rho_\phi$. The dark energy equation of state parameter may then be computed as
\begin{equation}
\label{wphi}
w_\phi  = -1-\frac{(\ln \rho_\phi)'}{3}\,.
\end{equation}

\section{Models with the same $w_\phi$ and $\Delta \alpha/\alpha$}

\subsection{Model I}

If $X$ is a small quantity, compared to the energy density associated with the scalar field potential, then a generic Lagrangian 
is expected to admit an expansion of the form
\begin{equation}\label{lagran}
{\mathcal L}_\phi=-V(\phi)+U(\phi)X+...\,,
\end{equation}
where $V$ and $U$ are real functions of $\phi$. For simplicity, here we shall consider that $V \ge 0$ and $U \ge 0$ and neglect terms of order two or higher in $X$ so that the Lagrangian in Eq. (\ref {lagran}) is simply given by
\begin{equation}\label{lagraneg}
{\mathcal L}_\phi=-V(\phi)+U(\phi)X\,.
\end{equation}
If $UX/V \ll 1$ then
\begin{equation}
w_\phi=-\frac{V-UX}{V+UX} \sim -1 + 2 \frac{UX}{V}\,.
\end{equation}
On the other hand $c_{s\phi}^2 = 1$.  Eq. (\ref {lagraneg}) can also be rewritten as 
\begin{equation}
{\mathcal L}_\phi={\mathcal L}_\psi = Y - V(\psi)\,,
\end{equation}
where $d\psi/d\phi=\pm {\sqrt U}$,  $Y=\psi_{,\mu} \psi^{,\mu}/2$ and $V(\psi) \equiv V(\phi(\psi))$. Consequently, 
\be 
\frac{\Delta\alpha}{\alpha}=\pm \beta \int  \frac{d \psi}{\sqrt {U(\psi)}}\,,
\ee
which implies that, in general, if $\Delta \alpha/\alpha$ is linear in $\phi$ then it must be non-linear in $\psi$. Consequently, the assumption that $\Delta \alpha/\alpha$ is linear in $\psi$ would lead to a biased estimation of the evolution of $\psi$ with redshift.

\subsection{Model II}

Another example comes from the tachyon Lagrangian
\be
{\cal L}_\phi=-U(\phi)\sqrt{1-2X}\,\,,
\ee
which has been proposed as a unified model for dark matter and dark energy (see \cite{Beca:2005gc,Avelino:2008zz,Avelino:2008cu} for a discussion of linear and non-linear aspects of unified dark energy models). In \cite{Avelino:2011ey} it has been shown that this model is dual, at the background level, to a comological model with both dark matter and a quintessence scalar field described by a standard Lagrangian
\be
{\cal L}_\psi=Y-V(\psi)\,,
\ee
with $d\psi/d\phi=\pm H(-3\dot\psi/2H_{,\psi})^{1/2}$, so that the background dynamics is the same for both models (here it is  implicitly assumed that $\dot\psi >0$ and $H_{,\psi} <0$ or vice versa).
Hence
\be 
\frac{\Delta\alpha}{\alpha}=\beta \phi=\pm \beta \int \left(\frac{-2H_{,\psi}}{3\dot\psi}\right)^{\frac12} \frac{d \psi}{H}\,.
\ee
which, analogously to the previous example, also implies that if $\Delta \alpha/\alpha$ is linear in $\phi$ then it must be non-linear in $\psi$, except if $H$ was constant. Again, the assumption that $\Delta \alpha/\alpha$ is linear in $\psi$ would bias the reconstruction of the dynamics of $\psi$.

In both models I and II we consider a correspondence (in the case of model II only at the background level) between cosmological models where time variations of fundamental couplings, such as $\alpha$ or $\mu$, are linearly coupled to a k-essence dark energy scalar field and a model where such variations are non-linearly coupled to a standard quintessence scalar field. Due to the non-linearity of the coupling, local Equivalence Principle constraints \cite{Olive:2001vz} on the value of $\beta$ do not necessarily apply at larger redshifts. This implies that the use of local (zero redshift) limits on the dark energy equation of state parameter or the time variation of fundamental "constants" to constrain the variation of fundamental couplings at higher redshifts is model dependent. Hence, many of the results presented in \cite{Avelino:2008dc,Dent:2008gx} relating local and cosmological variations of fundamental couplings, which apply to the standard case where the variation of fundamental parameters is driven by a linear coupling to a quintessence scalar field, need to be relaxed in the case a generic k-essence scalar field. 
\section{Models with the same $w_\phi$ and different $\Delta \alpha/\alpha$}

Consider the theory with the Lagrangian
\be\label{Lepsilon}
{\cal L}_\phi=X-V(\phi)+\epsilon\,X|X|\,,
\ee
where $\epsilon \ge 0$. Eq. (\ref{consphi}) can then be written as
\be\label{drho}
\rho^{\prime}_{\phi}=-3\phi_{,y}^2\left(\rho_{\phi}+\Omega_{m0}e^{-3y}+\epsilon\,\phi_{,y}^2\left(\rho_{\phi}+\Omega_{m0} e^{-3y}\right)^2\right)\,,
\ee
where $y=\ln{a}$.

If $\epsilon=0$, and considering the case with $\phi={\sqrt A}\,y$ where $0 \le A< 1$ is constant, one obtains
\be\label{rho0}
\rho_{\phi}=\frac{A\Omega_{m0}}{1-A}\,e^{-3y}+\left(1-\frac{\Omega_{m0}}{1-A}\right)e^{-3Ay}\,.
\ee
Using  Eq. (\ref{wphi}), the equation of state parameter becomes
\be\label{omega0}
w_\phi=-\frac{(1-A-\Omega_{m0})(1-A)}{A \Omega_{m0} e^{-3(1-A)y}+1-A-\Omega_{m0}}\,,
\ee
with 
\be
w_{\phi0}=\frac{A}{1-\Omega_{m0}}-1\,.
\ee

Here we shall assume that the evolution of $w_\phi$ is fixed so that Eqs. (\ref{rho0}) and  (\ref{omega0}) remain valid for any value of $\epsilon$. Hence, if $\epsilon \neq 0$ it is no longer true that $\phi_{,y}^2=A$ as in the $\epsilon=0$ case. Instead, one obtains 
\be
\phi_{,y}^2=\frac{-1+\sqrt{1+4\epsilon AH^2}}{2\epsilon H^2}\label{phiy}\,,
\ee
where $H^2$ is given by Eq. (\ref{fried1}) and $\rho_{\phi}$ is given by Eq. (\ref{rho0}). The value of $\phi$ may now be computed by numerically integrating Eq. (\ref{phiy}). If $\epsilon \ll 1$ then 
\be
\phi(y)={\sqrt A}\,\left(y+\frac{\epsilon}{6}\left(\rho_\phi+\Omega_{m0}-1\right)\right)\,,
\ee
up to first order in $\epsilon$. 

%%%%%%%%%%%%%%%%%%%%%%%%%%%%%%%%%%%%%%%%%%%%%%%%%%%%%%%%%%%%%%
\begin{figure}
\includegraphics[width=8cm]{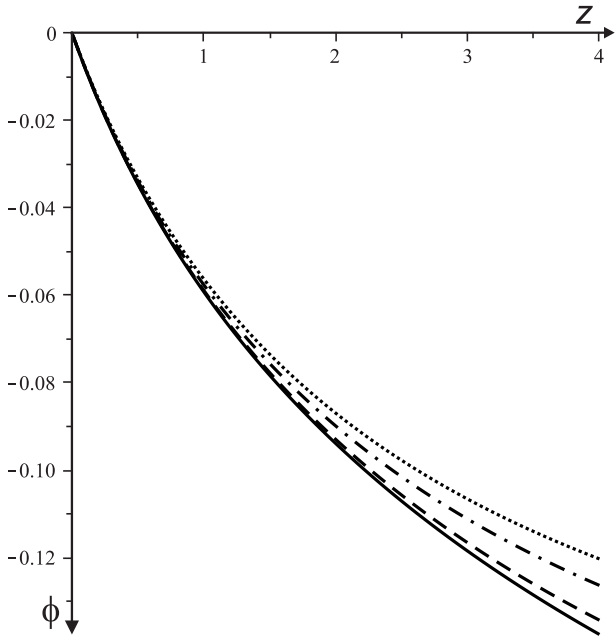}\caption{The evolution of the value of $\phi=\Delta \alpha/(\beta\alpha)$ as a function of redshift $z$ obtained using Eq. (\ref{phiy}) with various values of $\epsilon$: $\epsilon=0$ (solid curve),  $\epsilon=1$ (dashed curve), $\epsilon=5$ (dot-dashed curve) and $\epsilon=10$ (dotted curve).  $w_\phi(z)$ is the same for all models and is given by Eq. (\ref{omega0}).
\label{fig1}}
\end{figure}
%%%%%%%%%%%%%%%%%%%%%%%%%%%%%%%%%%%%%%%%%%%%%%%%%%%%%%%%%%%%%%

The left panel of Fig. 1 shows the evolution, as a function of the redshift $z=1/a-1$, of the value of $\phi=\Delta \alpha/(\beta\alpha)$ obtained using Eq. (\ref{phiy}) for $\epsilon=0$ (solid curve),  $\epsilon=1$ (dashed curve), $\epsilon=5$ (dot-dashed curve) and $\epsilon=10$ (dotted curve). It was assumed  that $w_{\phi0}=-0.99$, $\Omega_{m0}=0.27$ and $A=(w_{\phi0}+1)(1-\Omega_{m0})$. Fig. 1 shows large differences in the evolution of $\alpha$, despite the fact that the dark energy equation of state parameter $w_\phi(z)$ is the same for all models (given by Eq. (\ref{omega0}) and displayed in Fig. 2 (solid curve)). These results imply that the evolution of $w_\phi$ does not uniquely determine the evolution $\alpha$ (or other fundamental parameters), even in the case of a linear coupling.

\section{Models with the same $\Delta \alpha/\alpha$ and different $w_\phi$}

Consider again the Lagrangian given in Eq. (\ref{Lepsilon}) so that Eq. (\ref{drho}) describes the evolution of the energy density of the scalar field. At present the hints for a variation of $\alpha$ or $\mu$ with redshift remain controversial and, consequently, there is no strong reason to consider one possible variation over another. Let us assume that $\phi$ is a linear function of $y$ given by  $\phi(y)={\sqrt A} y$ with fixed $0 \le A < 1$. We shall use this function, containing only the first non-trivial term of the usual polynomial expansion of $\phi$ in powers of $y$, as an example of a model where the same variation of the fine structure constant with redshift may be associated with very different evolutions of the equation of state of dark energy. The energy density associated with the scalar field $\phi$ may be written as
\be
{\rho}_{\phi}=\tilde\rho_{\phi}+\epsilon A\,\rho_{\phi\epsilon}\,, \label{rhophi}
\ee
where $\tilde\rho_{\phi}$ is equal to the scalar field energy density given by Eq. (\ref{rho0}). The evolution of $\rho_{\phi \epsilon}$ may computed using Eq. (\ref{drho})  and is given by
\be\label{drhoper}
\rho^{\prime}_{\phi\epsilon}+3A\rho_{\phi\epsilon}=-3A \left(\rho_{\phi}+\Omega_{m0} e^{-3y}\right)^2\,,
\ee
up to zeroth order in $\epsilon A$. The solution is given by
\be
\rho_{\phi\epsilon}=C_1 e^{-3Ay}+C_2 e^{-6Ay}+C_3 e^{-3(A+1)y}+C_4 e^{-6y}\,, \label{rhophiepsilon}
\ee
with
\bq
C_1&=& \frac{2 \Omega_{m0}^2}{(A-2)}+2\Omega_{m0}-1\,,\\
C_2&=& \left(1-\frac{\Omega_{m0}}{1-A}\right)^2\,,\\
C_3&=& \frac{2A \Omega_{m0}}{(1-A)}\left(1-\frac{\Omega_{m0}}{1-A}\right)\,\\
C_4&=& \frac{A \Omega_{m0}^2}{(1-A)^2(2-A)}\,.
\eq
Here, $C_1$ is such that the condition $\rho_{\phi\epsilon0}=0$ is verified (so that ${\rho}_{\phi 0}=\tilde\rho_{\phi 0}=\Omega_{\phi 0}$). We have verified numerically that, for $\epsilon A \ll 1$, Eqs. (\ref{rhophi}) and (\ref{rhophiepsilon}) provide an excellent approximation to the true result, at least while $w_\phi$ remains smaller than zero.

%%%%%%%%%%%%%%%%%%%%%%%%%%%%%%%%%%%%%%%%%%%%%%%%%%%%%%%%%%%%%%
\begin{figure}
\includegraphics[width=8cm]{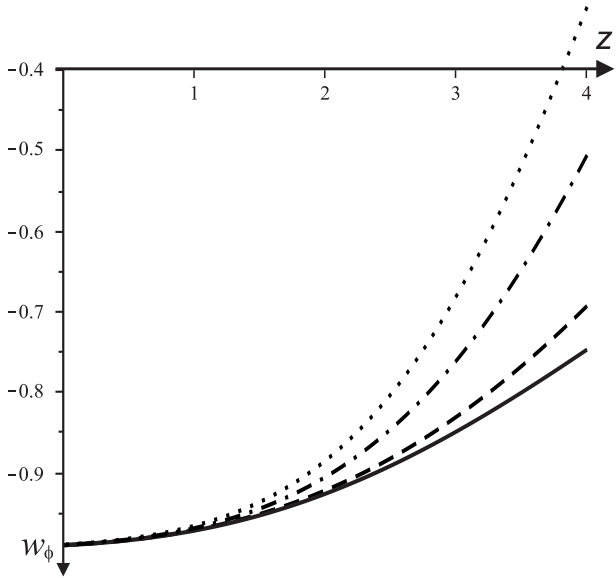}\caption{The evolution $w_{\phi}$ with redshift $z$ obtained using Eq. (\ref{wphi1}) with various values of $\epsilon$: $\epsilon=0$ (solid curve),  $\epsilon=1$ (dashed curve), $\epsilon=5$ (dot-dashed curve) and $\epsilon=10$ (dotted curve). The evolution of $\phi=\Delta \alpha/(\beta\alpha)$ with $z$ is the same for all models.
\label{fig2}}
\end{figure}
%%%%%%%%%%%%%%%%%%%%%%%%%%%%%%%%%%%%%%%%%%%%%%%%%%%%%%%%%%%%%%

The corresponding equation of state has the form
\be
w_\phi={\tilde w}_\phi+\epsilon A\, w_{\phi\epsilon}\label{wphi1}\,,
\ee
where ${\tilde w}_\phi$ is the equation of state parameter given by Eq. (\ref{omega0}) and 
\be
w_{\phi \epsilon}=-\frac13\left(\frac{\rho_{\phi \epsilon}^\prime}{{\tilde \rho}_{\phi}}-\frac{{\tilde \rho}_\phi^\prime\rho_{\phi\epsilon}}{{\tilde \rho}_{\phi}^2}\right)\,.
\ee
Hence,
\be
w_{\phi 0}={\tilde w}_{\phi 0}+\epsilon A\, w_{\phi\epsilon 0}={\tilde w}_{\phi 0}+\epsilon\left({\tilde w}_{\phi 0}+1\right)^2 (1-\Omega_{m0})\label{wphi0}\,,
\ee
so that $w_{\phi 0}\sim{\tilde w}_{\phi 0}$  (note that for $\epsilon=10$, $\Omega_{m0}=0.27$ and ${\tilde w}_{\phi 0}=-0.99$ one has $w_{\phi 0}-{\tilde w}_{\phi 0} = 7 \times 10^{-4}$). We have verified both analytically and numerically that, for $\epsilon A \ll 1$, the results obtained considering that $\phi(y)={\sqrt A} y$ (with fixed $A$, so that $w_{\phi 0}$ is given by Eq. (\ref{wphi0})) are an excellent approximation to those which one would find if $w_{\phi 0}$ had been fixed exactly.

Fig. 2 shows the evolution $w_{\phi}$ with redshift $z$ obtained using  Eq. (\ref{wphi1}) with $\epsilon=0$ (solid curve),  $\epsilon=1$ (dashed curve), $\epsilon=5$ (dot-dashed curve) and $\epsilon=10$ (dotted curve). Again, it was assumed that  $w_{\phi0}=-0.99$, $\Omega_{m0}=0.27$ and $A=(w_{\phi0}+1)(1-\Omega_{m0})$. Fig. 2 shows that the same evolution of $\alpha$ may be consistent with many different dynamics of $w_\phi$, depending on the k-essence model which describes the dynamics of the dark energy scalar field. Again, this is true even in the case, considered in the present letter, where $\phi$ is linearly coupled to $\Delta \alpha/\alpha$.

\section{\label{conc}Conclusions}

Despite the good prospects for a significant improvement of the constraints on the variation of $\alpha$ and $\mu$ in the coming years, particularly with forthcoming data to be obtained with the ESPRESSO and CODEX spectrographs, respectively for the VLT and E-ELT \cite{Cristiani:2007by}, the relevance of a future unambiguous determination of such variation for the reconstruction of the dark energy dynamics depends crucially on the dark energy being described by a standard quintessence scalar field linearly coupled to the variation of $\alpha$ and $\mu$. In this letter, we relaxed this assumption and considered dark energy models where the dark energy role is played by a more generic k-essence scalar field. We have shown that, in general, there is not a one-to-one correspondence between the evolution of the dark energy equation of state and the evolution of varying fundamental couplings. This is true even if the evolution of the dark energy scalar field is assumed to be linearly coupled to varying fundamental couplings such as $\alpha$ and $\mu$. Hence, additional knowledge about the scalar field lagrangian describing the dynamics of dark energy is required for a successful reconstruction of the equation of state of dark energy using varying couplings.  It is crucial that this is  taken into account in any future attempt of dark energy reconstruction using varying couplings.

In particular, if the evolution of $\alpha$ and $\mu$ is confirmed unambiguously by future data then it needs to be consistent with the very stringent low redshift bounds \cite{Peik:2006xy,Rosenband:2008,Gould:2006,Petrov:2006,Onegin:2010kq}. In this case, an important test to the usual assumption of a canonical kinetic term with a linear coupling (see, for example, \cite{Dent:2009}) could be made, for redshifts in the range $z=1-4$, by verifying whether or not the data turns out to be consistent with a null evolution with redshift of the ratio $\alpha/\mu$. Also, a comparison (at relatively low redshifts) between the inferred dark energy dynamics using varying couplings and using standard methods, such as type Ia supernovae, galaxy clustering or weak lensing, should be possible with future data from the Euclid Mission. Furthermore, although it has been demonstrated that if the combined dynamics of dark energy and varying couplings is described by a standard quintessence field then the spatial variations of the couplings would be negligible \cite{Barrow:2005sv,Shaw:2005gt,Avelino:2005pw,Avelino:2008cu}, a large sample of quasar absorption line spectra obtained using Keck telescope observations and new Very Large Telescope data has now been shown to be consistent with a dipolar spatial variation of the fine structure constant \cite{Webb:2011,King:2012}. Hence, if the interpretation of these results as evidence for the spatial variation of $\alpha$ turns out to be true then the models connecting the dynamics of dark energy to the evolution of fundamental parameters of nature will need to be further revised to take into account, for example, the role of cosmic domain walls in seeding spatial fluctuations of fundamental couplings such as $\alpha$ or $\mu$ \cite{Avelino2001148,Menezes:2004tp,Olive:2010vh,Chiba:2011en,Bamba:2011nm,Olive:2012ck}.

%%%%%%%%%%%%%%%%%%%%%%%%%%%%%%%%%%%%%%%%%%%%%%%%%%%%%
\begin{acknowledgments}

This work is partially supported by FCT-Portugal through the project CERN/FP/116358/2010. We also thank CNPq for financial support.

\end{acknowledgments}

%%%%%%%%%%%%%%%%%%%%%%%%%%%%%%%%%%%%%%%%%%%%%%%%%%%%%%%%%%

\bibliography{alpha}

\end{document}